# Low paleopressure of the Martian atmosphere estimated from the size distribution of ancient craters

Edwin S. Kite[1,2,3], Jean-Pierre Williams[4], Antoine Lucas[1], and Oded Aharonson[1,5]

[1] *Geological and Planetary Sciences, Caltech, Pasadena, California 91125, USA.*

[2] *Department of Astrophysical Sciences, Princeton University, Princeton, New Jersey 08544, USA.*

[3] *Department of Geosciences, Princeton University, Princeton, New Jersey 08544, USA.*

[4] *Department of Earth, Planetary and Space Sciences, University of California - Los Angeles, Los Angeles, California 90095, USA.*

[5] *Helen Kimmel Center for Planetary Science, Weizmann Institute of Science, Rehovot 76100, Israel.*

**The decay of the martian atmosphere – which is dominated by carbon dioxide – is a component of the long-term environmental change on Mars[1] from a climate that once allowed rivers to flow to the cold and dry conditions of today[2-6]. The minimum size of craters serves as a proxy for palaeopressure of planetary atmospheres, because thinner atmospheres permit smaller objects to reach the surface at high velocities and form craters[7–9]. The Aeolis Dorsa region near Gale crater on Mars contains a high density of preserved ancient craters interbedded with river deposits[11] and thus can provide constraints on atmospheric density around the time of fluvial activity. Here we use high-resolution orthophotos and digital terrain models[10] from the Mars Reconnaissance Orbiter to identify ancient craters in Aeolis Dorsa that date to about 3.6 Gyr ago and compare their size distribution with models of atmospheric filtering of impactors[12,13]. We obtain an upper limit of 0.9±0.1 bar, rising to 1.9±0.2 bar if rimmed circular mesas – interpreted to be erosionally-resistant fills of floors of impact craters – are excluded. We assume target properties appropriate for desert alluvium[14]: if sediment had rock-like rock-mass strength**

**similar to bedrock at the time of impact, the upper limit increases by a factor of up to two. If Mars did not have a stable multibar atmosphere at the time that the rivers were flowing – as suggested by our results – then the warm and wet $CO_2/H_2O$ greenhouse of Ref. 2 is ruled out, and long-term average temperatures were most likely below freezing.**

Planetary atmospheres brake, ablate, and fragment small asteroids and comets, filtering out small high-velocity surface impacts and causing fireballs, airblasts, meteors, and meteorites. The smallest impact craters near sea-level on Earth have diameter $D \sim 20$ m. "Zap pits" as small as 30 μm are known from the airless Moon, but other worlds show the effects of progressively thicker atmospheres: the modern Martian atmosphere can remove >90% of the kinetic energy of >240 kg impactors[7]; Titan's paucity of small craters is consistent with atmospheric filtering of craters smaller than 6-8 km (Ref. 8); and on Venus, craters $D < 20$ km are substantially depleted by atmospheric effects[9].

Changes in the concentration of atmospheric volatiles are believed to be the single most important control on Mars climate evolution and habitability, which in turn is a benchmark for habitable-zone calculations for exoplanets[15]. Contrary to early work[2], it is doubtful that increasing $CO_2$ pressure (≈total atmospheric pressure, $P$) is enough to raise early Mars mean-annual surface temperature ($\overline{T}$) to the freezing point, even when water vapor and cloud feedbacks are considered[5]. However, increased $CO_2$ aids transient surface liquid water production by impacts, volcanism, or infrequent orbital conditions[3-4,6]. Existing data requires an early epoch of massive atmospheric loss to space, suggests that the present-day rate of escape to space is small, and offers evidence for only limited carbonate formation[16]. These data have not led to convergence among atmosphere evolution models, which must balance poorly understood fluxes from volcanic degassing, escape to space, weathering, and photolysis[17]. More direct measurements[18] are required to determine the history of Mars' atmosphere. Wind erosion exposes ancient cratered volumes on Mars, and the size of exhumed craters has been previously suggested as a proxy of ancient Mars $P$ (e.g., Ref 19).

Here we obtain a new upper limit on early Mars atmospheric pressure from the size-frequency distribution of small ancient craters interspersed with river deposits in Aeolis, validated using High Resolution Imaging Science Experiment (HiRISE) DTMs and anaglyphs, in combination with simulations of the effect of $P$ on the crater flux. The craters are interbedded with river deposits up to ~$10^3$ km long, with inferred peak river discharge 10-1000 m$^3$/s (Ref. 11). Therefore, the atmospheric state they record corresponds to an interval of time when Mars was substantially wetter than the present, probably > 3.6 Ga (Supplementary Material).

Aeolis Dorsa (part of the Medusae Fossae Formation) is a promising location to hunt for ancient craters: the stratigraphy contains numerous channel-fill deposits of large rivers, and when these overlie a crater, that crater must be as ancient as the rivers[20]. Certain beds in Aeolis Dorsa (Supplementary Material) preserve a high density of ancient craters, perhaps due to slow deposition or a diagenetic history unusually favorable for crater preservation. We constructed stereo DTMs/orthophotos for two image-pairs covering these beds, DTM1 and DTM2 (Methods). Following a checklist (Supplementary Table), craters were classified as definite ancient craters (visibly embedded within stratigraphy: e.g., overlain by river deposit) ($n = 56$, median diameter $D_{50} = 107$ m, 10$^{th}$-percentile diameter $D_{10} = 50$ m), rimmed circular mesas (RCM) ($n = 71$, $D_{50} = 48$m, $D_{10} = 21$m), or candidate ancient crater ($n = 192$, $D_{50}$ also 48m, $D_{10}$ also 21m; candidates are not considered further, but their inclusion would strengthen our conclusions). We measured $D$ by fitting circles to preserved edges/rims. RCM appear as disks in raw HiRISE images. We interpret them as the erosionally-resistant fills/floors of impact craters that were topographically inverted during the deflation of the target unit. They are unlikely to be outliers of a young mantle because they are not found away from the fluvial unit. We plot them separately, but consider them to be probable ancient craters. We used unambiguously ancient craters as a guide to the preservation state of the smaller craters. These ancient craters are unlikely to be maars; maars are not randomly distributed in space or time/stratigraphy. We also reject the possibility that they are paleo-karst sinkholes; sinkholes lack rims, are concentrated at particular stratigraphic levels, and are overdispersed.

We generated synthetic crater populations for varying $P$ (ref. 12). The approach is conceptually similar to that of previous studies[13], and benefits from measurements of the current Martian cratering flux (Methods, Supplementary Material). Modeled smallest-crater diameter increases linearly with pressure (~20 m at 1 bar) as expected from equating impactor and atmospheric-column masses (Melosh, 1989). This is broadly consistent with low-elevation impacts on Earth (the column mass of Earth's sea-level atmosphere is equivalent to ~ 0.4 bar on Mars). We apply a geometric correction for exhumation from a cratered volume (Supplementary Material) assuming that initial crater shape is isometric over the diameter range. After bayesian fitting, we correct our $P$ estimate for elevation (our DTMs are below datum; Mars-average $P$ was 20% lower than local $P$).

We compared the model to the combined dataset (DTM1+DTM2). Combined best fits are $P = 1.9\pm0.2$ bar, falling to $P = 0.9\pm0.1$ bar if RCM (candidate syndepositional impact craters) are also included (Figure 2). Because better preservation/exposure could allow still smaller embedded craters to be uncovered, we interpret our fits as upper limits. The best fit to DTM1 (DTM2) ancient craters alone is 1.7±0.3 bar (2.2±0.3 bar), falling to 0.8±0.1 bar (0.9±0.1 mbar) if RCM are included.

The results are sensitive to target strength, as expected[21]. Increasing the target rock-mass strength to a hard-rock-like 6.9 MPa (ref. 22) while holding all other parameters constant increases the combined upper limit on $P$ to ~2 bar (Supplementar Material). Our work assumes weak soil-like target strength appropriate for river alluvium in an aggrading sedimentary deposit: if sediment developed bedrock-like rock-mass strength by early diagenesis, the upper limit is greatly increased. Sensitivity tests show a relatively minor effect of fragmentation on the results (Supplementary Material).

We do not consider crater shrinkage or expansion by crater degradation. Only shrinkage matters for the purpose of setting an upper bound on *P*: as the crater is abraded, the exposed radius must eventually vanish. We surmise that shrinkage is a small effect because impact craters are bowl-shaped (as opposed to cone-shaped), and because rims are frequently preserved.

Our technique rules out a thick *stable* paleoatmosphere, and cannot exclude atmospheric collapse-reinflation cycles on timescales much shorter than the sedimentary basin-filling time. General Circulation Model (GCMs) predict that atmospheric collapse to form $CO_2$-ice sheets and subsequent reinflation might be triggered by obliquity change[5]. If sediment accumulated at 1-100 µm/yr (ref. 20), our DTMs could integrate over ~$10^6$-$10^8$ years of sedimentation and contain many collapse-and-reinflation cycles. Therefore one interpretation is that smaller ancient craters formed while the atmosphere was collapsed, while rivers formed during high-obliquity, thick-atmosphere intervals. However, published models indicate that collapse to form polar $CO_2$-ice sheets only occurs for pressures less than our upper limit.[5] If these models are correct, then our pressure constraint is a true upper bound on typical atmospheric pressure.

Downward revisions to $CO_2$'s infrared opacity indicate that *any* amount of $CO_2$ is insufficient to warm early Mars $\overline{T}$ to the freezing point[5]. Even if further work incorporating radiatively-active clouds[23] moderates this conclusion, our result is an independent constraint on stable $CO_2$/$H_2O$ warm-wet solutions (Figure 3). However, increased $CO_2$ below the warm-wet threshold primes Mars climate for surface liquid water production by other relatively short-lived mechanisms, by adding to the greenhouse effect, pressure-broadening the absorption lines of other gases[4], suppressing evaporitic cooling[6], and increasing atmospheric heat capacity[3].

If the small-crater limit is representative of early Mars $P$, it is difficult to envisage continuous stability of surface liquid water for the $10^4$-$10^5$ yr needed to allow water to cycle between deep aquifers and the surface. This is true even with optimistic $CO_2$ radiative-forcing parameterizations. Transient warming by eruptions, impacts, or infrequent orbital conditions could unfreeze the surface and shallow subsurface, allowing runoff, but would not last long enough to unfreeze ground at ~1 km depth. Therefore, $CO_2$/$H_2O$ atmospheric models do not support $\overline{T}$ above freezing on early Mars, which has implications for sedimentary-rock formation and diagenesis, groundwater hydrology, and habitability. (A new study[24] shows that mean temperatures above the freezing point are marginally consistent with our result, but only if the early Mars atmosphere contained ≥20% $H_2$).

Atmospheric loss must be part of the explanation for Mars' great drying, if only because freshwater rivers cannot flow for hundreds of km when simultaneously boiling and freezing. How high $P$ was, and its decay over time, are not known. The 2014-2015 MAVEN mission will measure modern loss processes, which is complementary to our geologic paleo-proxy approach.

Mars would have formed with ≥6-10 bars $CO_2$-equivalent of carbon assuming the same initial [C] and [Kr] as Earth. $^{40}Ar/^{36}Ar$ and $^{129}Xe/^{132}Xe$ suggest that 90-99% of the initial atmosphere was lost prior to ~4.1 Ga[25]. Subsequent loss rates are less clear; Mars' $C/^{84}Kr$ ratio suggests $P$ ~ 60 mbar following the Late Heavy Bombardment[25].

Ref. 18 uses a volcanic bomb sag in Gusev crater to infer $P > 120$ mbar from the bomb sag's terminal velocity. This is consistent with our result. Our small-crater constraints on early Mars atmospheric pressure are also congruent with isotopic and mineralogic indicators, which generally require more assumptions than our method. For example, prehnite is observed on Mars and is unstable for $CO_2$ mixing ratios $>2 \times 10^{-3}$. This implies $P \lesssim 1$ bar, but only if water at depth was in equilibrium with the atmosphere. The composition of a carbonate-rich outcrop at Gusev has been interpreted to require $P = 0.5 – 2$ bar assuming that the carbonates are a solid solution in thermodynamic equilibrium[26]. Models[16] of volcanic degassing, impact delivery, and escape of $CO_2$ also hint that Mars atmospheric pressure at the time of the Late Heavy Bombardment was not greater than our estimate.

In the future, pyroclastic-blast runout length or even rainsplash[27] could be used to constrain *P*. Curiosity's field site in Gale crater contains syndepositional craters (Figure 1a), so Curiosity could validate orbital identifications of embedded craters along its traverse. The 40-year-old prediction of a connection between drying and atmospheric decay could be tested by applying the small-crater technique to sedimentary deposits of different ages – ranging from Mawrth (the oldest known sedimentary sequence in the Solar System), through Meridiani, to the relatively young Valles Marineris silica deposits. This could yield a time series of constraints on early Mars atmospheric pressure, stratigraphically coordinated to the sedimentary record of Mars' great drying.

**Methods.**

**DTM generation.** DTMs were constructed following Ref. 10 from HiRISE images PSP_007474_1745/ESP_024497_1745 (DTM 1) and ESP_017548_1740/ESP_019104_1740 (DTM 2). MOLA Precision Experiment Data Records (PEDRs) were used as ground control points. Optimal resolution depends on HiRISE image map scale (0.25-0.5m), giving 1 m/post – 2.5 m/post DTMs. Vertical precision is ~0.3 m, with 90% probability of precision ≲1 m (Supplementary Material).

**Cratering model.** We build a synthetic impactor population by drawing randomly from a size distribution constrained by satellite observations, and an estimated initial-velocity distribution of meteoroids at Mars' orbit. Distributions of material types, densities, and ablation coefficients $k_{ab}$ are set based on terrestrial fireball network observations (the model assumes the same fractional distribution at Mars). Details of the cratering model, including the sources for parameter choices, are given in the Supplementary Material. We advect these populations to the surface through atmospheres with scale height 10.7 km. The code does not track planet curvature; we do not allow impactors to skip back to space. The atmosphere drains kinetic energy from impactors via drag (per unit mass),

$$dv/dt = C_D \, \varrho_a \, v^2 \, A \qquad (1)$$

(we assume a drag coefficient $C_D = 1$ across the velocity ($v$) and size range of interest)

and ablation,

$$dm/dt = (C_h \, \varrho_a \, v^3 \, A \,) / 2\zeta \qquad (2)$$

where $\varrho_a$ is local atmospheric density, $A$ is cross-sectional area, $C_h$ is the heat transfer coefficient, and $\zeta$ is the heat of ablation. Particles braked to <500 m/s would not form hypervelocity craters and are removed from the simulation. We do not track secondary craters, because meter-sized endoatmospheric projectiles are likely to be braked to low speeds for the relatively thick atmospheres we are evaluating. In other words, if wet-era small craters are secondaries, then early Mars' atmosphere was thin. Transient impact-induced increases in $P$ would not affect our upper limit; transient local decreases in $P$ could conceivably enhance secondary flux. It has been suggested that unrecognized secondary craters significantly contribute to all counts of $D < 1$ km Martian craters[28], and although it has been shown by others that unrecognized secondaries are not required to explain observed crater size-frequency distributions[12,29-30], the small-crater-rich size-frequency distribution of secondaries (if pervasively present with high relative frequency on all Martian surfaces, as suggested by Ref. 28) could mimic the effect of a lower atmospheric pressure. Crater sizes are calculated using π-group scaling[22], assuming a target strength of 65 kPa and a target density of 2000 kg/m³ appropriate to cohesive desert alluvium[14], with the Holsapple scaling parameters $k_1$=0.132, $k_1$ = 0.26, $k_r$ = 1.1, and $\mu$=0.41. We adopt the values in

`keith.aa.washington.edu/craterdata/scaling/theory.pdf`; note that the value $k_1$ = 0.24 given in Table 1 of Ref. 22 is in error (K.A. Holsapple, pers. comm.).

**Author contributions**

**Corresponding author.** Correspondence to Edwin Kite (ekite@caltech.edu).

**Acknowledgements.** We thank Ingrid Daubar, Joe Dufek, Bethany Ehlmann, Woody Fischer, Vamsi Ganti, Itay Halevy, Kevin Lewis, Michael Manga, Ramses Ramirez, Melissa Rice, and Alejandro Soto for preprints and discussions. We thank the HiRISE team, the CTX team, our reviewers, and the Associate Editor. This work was funded by an O.K. Earl Fellowship (to E.S.K.), and by the U.S. taxpayer via NASA grants NNX11AF51G (to O.A.) and NNX11AQ64G (to J.-P.W.)

**Figures.**

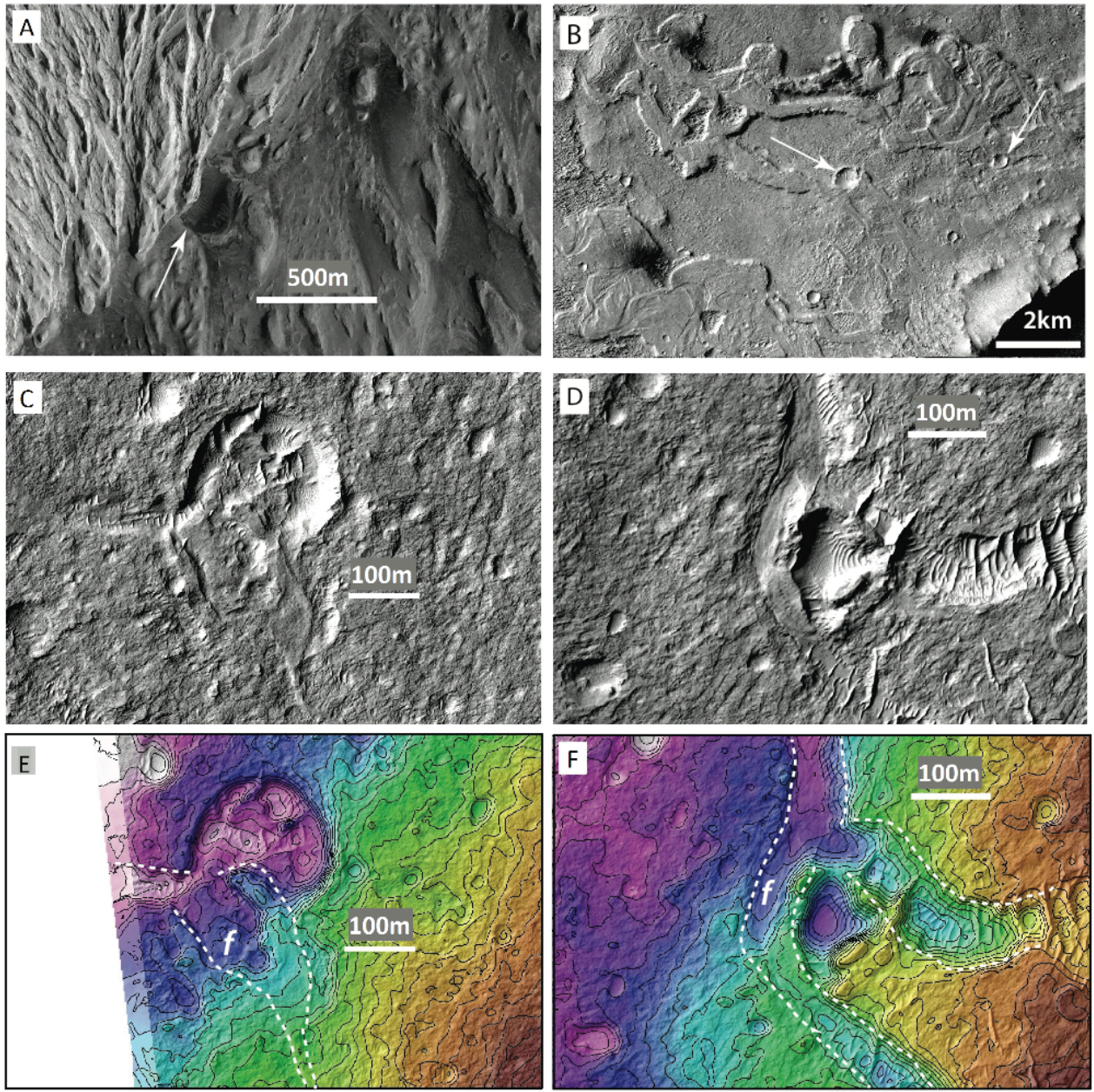

**Figure 1. Gallery of ancient Martian craters** (after Ref. 20). a) Crater being exhumed from beneath an unconformity within Gale crater's mound (Aeolis Mons/Mt. Sharp), the destination of the Curiosity rover. ESP_019988_1750. b) Craters with intact rims being exhumed from beneath

meander belts, Aeolis Dorsa, G03_019249_1744_XI_05S205W; Ref. 11. c) Crater partly draped by fluvial channel materials (*f*), Aeolis Dorsa. 238 m diameter. ESP_019104_1740. d) Crater partly draped by fluvial channel materials *f*, Aeolis Dorsa, ESP_019104_1740. 141m diameter. e) Crater from (c), but with 1m elevation contours from DTM2 (see text). f) Crater from (d), with 1m contours from DTM2.

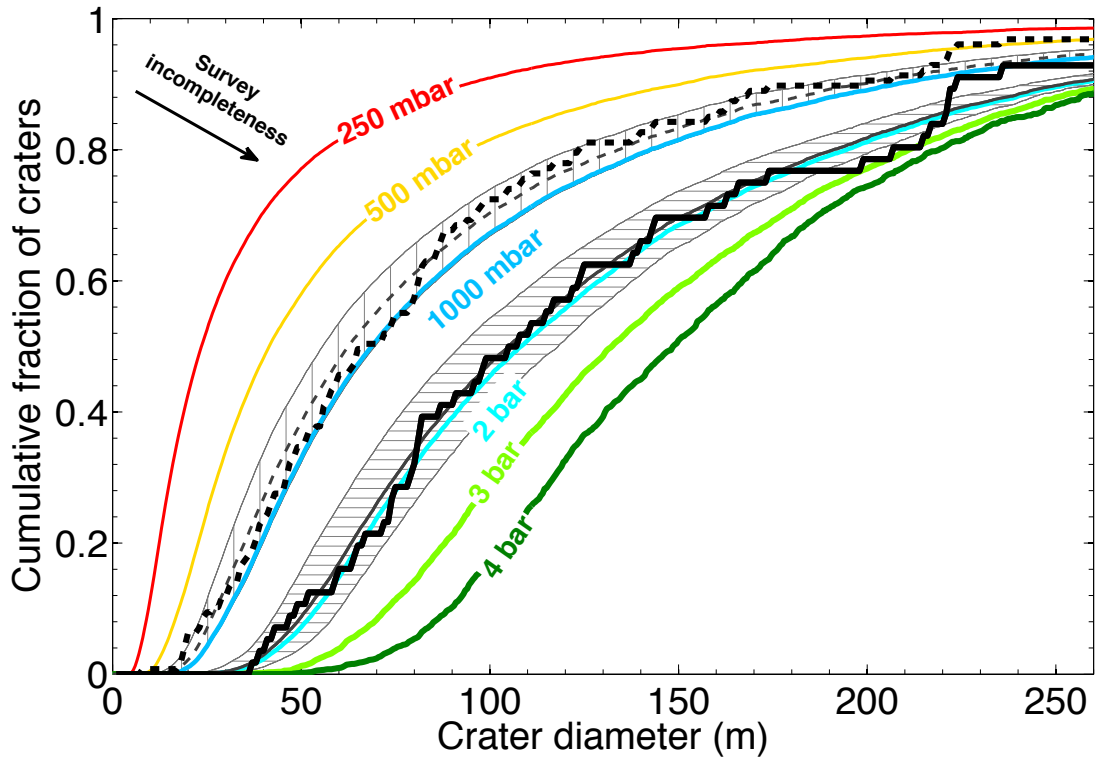

**Figure 2. Upper limits on Early Mars atmospheric pressure:** comparison of model crater size-frequency distributions to observations. Solid black line corresponds to definite embedded craters. Dashed black line additionally includes rimmed circular mesas. Stair-stepping in the data curves corresponds to individual craters. Colored lines show model predictions for atmospheric filtering of small impactors at different pressures. Gray hachured regions correspond to 2σ statistical-error envelope around the best-fit paleopressure to the data (best fits shown by thick

gray lines). Survey incompleteness leads to overestimates of median crater size, so best fits are upper limits.

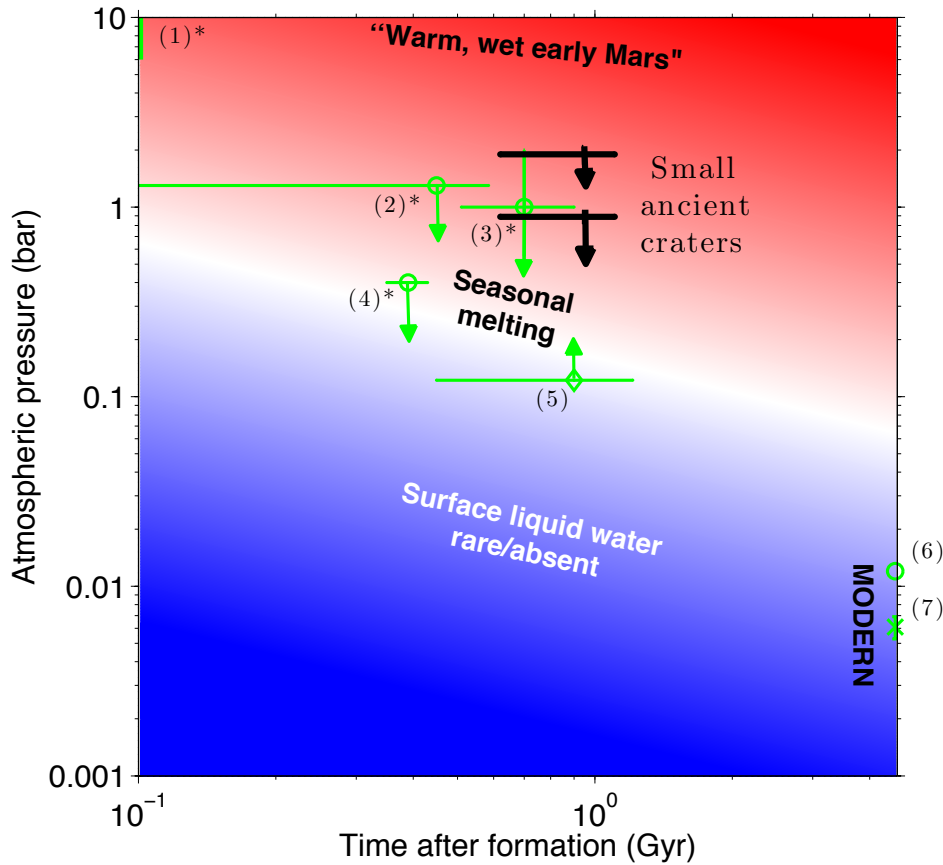

**Figure 3**. **Paleopressure constraints on the great drying of Mars.** Black symbols are the result from this work. Green symbols are other estimates as follows (asterisks mark indirect estimates):- (1*) cosmochemical estimate[1]; (2*) prehnite stability; (3*) carbonate Mg/Ca/Fe (ref. 25); (4*) $^{40}Ar/\,^{36}Ar$ (ref. 31); (5) bomb sag[18]; (6) modern atmosphere; (7) modern atmosphere + $CO_2$ ice. Approximate and model-dependent implications for sustained surface habitability are shown by background colors (blue = always below freezing, red = melting year-round, slope schematically shows the effect of the fainter young sun). "Warm, wet early Mars" refers to Ref. 2's stable climate solution. Age estimates are detailed in Supplementary Material.

**Supplementary Material.**

**Contents: Supplementary Text, Supplementary Table 1, Supplementary Figures 1-8.**

**1. Geologic Constraints and Geologic Context.**

**1a. Stratigraphic Control.**

Stratigraphic relations prove that our DTMs sample near the center of a thick interval of fluvial deposition; therefore, the rivers in our study area do not represent the final gasp of large-river activity. The most recent published map covering Aeolis Dorsa is Zimbelman & Scheidt (2012). Our DTMs straddle the contact of two fluvial units (Fig. S1) within the area mapped by Zimbelman & Scheidt as "AHml1." These units are traceable for >300 km. The lower of the two units, which we informally term F1 (Fluvial 1), contains broad meander-belts. Material laterally adjacent to channel belts erodes to form yardangs, leaving the meander-belts as locally high-standing features. F1 is overlain, apparently conformably, by F2 (Fluvial 2). The surface trace of this contact intersects both of our DTMs. F2 is a slope-forming, smoothly-eroding unit, densely peppered with rimless craters, interpreted as impact craters. Across Aeolis Dorsa, F2's observed crater density is higher than that of the units which sandwich it, especially near the contact with F1. F2 is associated with young aeolian bedforms. We interpret the sediment source for these bedforms to be erosion of F2. The erosional expression of channels in F2 is variable, but relative to channels in F1 they are typically narrower, have more frequent confluences, form more tree-like as opposed to subparallel networks, and are less frequently preserved in inverted relief than are channels in F1. F2 is >100m thick and is overlain by additional channel-containing units (not obviously exposed in our DTMs) that feature channel belts wider than those in F2. In all cases, channels show little relationship to the modern topography (e.g. Lefort et al., 2012) and the channels are eroding out of the rock. Because the channels are embedded in the stratigraphy, F2 channels postdate F1 channels. The base of F1 is not exposed near our study region, but it is at least tens of meters below the F1-F2 contact. Because our DTMs sample at/near the base of a thick channel-containing unit that is overlain by further channel-containing units, we conclude that our *P* constraint corresponds to the heart of a major river-forming time interval on Mars (conceivably, the *only* major river-forming time interval on Mars; Howard et al., 2005). The total stratigraphic interval over which fluvial deposits are abundant in Aeolis Dorsa is >300m.

The simplest interpretation of the interfluve materials in both F1 and F2 is that they consist of the overbank deposits of rivers, but other interpretations are possible. For example, the river deposits could be the fill of incised valleys that postdate the interfluve materials.

**1b. Age Control**

The craters date from around the time when large rivers flowed on the surface of Mars; they are almost certainly pre-Amazonian, and probably Early Hesperian or older. We carried out a CTX crater count over an 8.3 x $10^4$ km$^2$ region largely conterminous with Aeolis Dorsa (Fig. S2a), categorizing craters > 1km in diameter as 'postfluvial,' 'synfluvial/prefluvial,' and 'undetermined' on the basis of local crosscutting relationships. Based on crater morphology we think most of the 'undetermined' craters are in fact postfluvial, implying a N(1) Crater-Retention Age (CRA) on the Hesperian/Amazonian boundary and an N(2) CRA straddling the Late Hesperian/Early Hesperian boundary (where N(x) is the frequency of craters with $D$ > x km per $10^6$ km$^2$ count area; Werner & Tanaka, 2011) (Fig. S2b). Stratigraphic relations (Zimbelman & Scheidt, 2012), buttes that we interpret as outliers of formerly sheet-like stratigraphic units, and the shallower slopes of the diameter-frequency curves (Smith et al., 2008) for craters <2km diameter (Fig. S2b) all strongly suggest removal of several hundreds of meters of overburden. Removal of overburden would also remove craters, so our CRAs are minima. This further supports our inference that the rivers flowed in the Hesperian or Late Noachian. Excluding craters <2 km diameter for which overburden-removal effects are most severe, the nominal ages from `craterstats2` (Michael and Neukum, 2010) fits to these data are 3.44 (+0.06/–0.10) Ga for the postfluvial population ($n$ = 34; red triangles in Figure 2b), 3.61 (+0.03/–0.04) Ga additionally including the undetermined population (total $n$ = 52; blue circles in Figure 2b), and 3.71 (+0.02/–0.03) Ga additionally including synfluvial/prefluvial craters (total $n$ = 68; green squares in Figure 2b). These nominal ages adopt the Ivanov (2001) production function (PF) and the Hartmann & Neukum (2001) chronology function (CF).

Our preferred nominal age for the rivers (postfluvial craters + undetermined craters) is identical to the formation age of Gale crater reported by Le Deit et al. 2013 using the same PF and CF (3.61 (+0.04/–0.06) Ga). This suggests that our paleopressure constraint applies to the sedimentary deposits infilling Gale crater, reinflating a thin atmosphere via post-Noachian volcanic degassing is difficult (Stanley et al. 2011).

Our DTMs lie within a region of Aeolis Dorsa (Figure S2a) that has an unusually low N(1): if this results from relatively rapid exhumation, consistent with the excellent preservation state of the ancient river deposits, a resurfacing rate of ~1 μm/yr is implied over $10^{8-9}$ yr timescales. Relatively rapid modern erosion, combined with a high embedded-crater density, makes this a particularly attractive site for our procedure. Rapid erosion minimizes the proportion of geologically-recent (synerosional) craters in the crater population, and thus the impact of false positives (assuming that the fraction of young craters falsely classified as ancient is fixed). Our results are consistent with Zimbelman & Scheidt (2012), who additionally suggest that the rivers (i.e. Zimbelman & Scheidt's "AHml1") predate a topographically high-standing unit (their "Hmm," surrounding Asau crater) with a ~3.7 Ga CRA on the Hartmann & Neukum (2001) chronology. Regional geology as mapped by Irwin & Watters (2010) implies that the rivers are not older than Late Noachian.

We briefly explain the chronological constraints shown for the other data points in Fig. 3. The prehnite ("2*") age estimate assumes prehnite formation prior to the Isidis impact (Fassett & Head, 2011), consistent with although not required by geologic relations (Ehlmann et al. 2011); the carbonate Mg/Ca/Fe ("3*") age estimate assumes that the Comanche outcrop formed after the Gusev impact but prior to the Gusev plains lavas (Greeley et al., 2005); for the $^{40}Ar/^{36}Ar$ age constraint ("4*") we use the 4.16±0.04 Ga age adopted by Ref. 31; and for the bomb sag ("5*") age estimate we assume a pre-Amazonian age. All of these ages – with the possible exception of the ALH 84001 age – may need later revision; the crater chronology of early Mars has not yet been securely calibrated to radiogenic dates (Robbins et al., 2013).

**2. Details of Small Crater Analysis.**

When craters are dispersed through a 3D volume (Edgett & Malin, 2002), the size-frequency distribution of craters exposed at the surface will favor larger craters. This is because a 2D surface cutting through the volume (e.g., the erosion surface) is more likely to intersect a big crater than a small one. This geometric exposure correction is proportional to crater size if craters of different sizes have the same shape. This is approximately true in the strength regime relevant to this paper (Melosh, 1989). If craters of different sizes have the same shape, then crater area is proportional to the square of diameter, but the probability of a plane cutting through a crater is proportional to diameter. Therefore, we apply a correction proportional to crater size.

In Aeolis Dorsa, sediment moved by small impact events is a small fraction of the total sediment moved by all erosion and sedimentation processes. Therefore, in Aeolis Dorsa, small craters can be thought of as tracer particles with respect to erosion and sedimentation processes. Scale-independence of erosion and sedimentation events (the Sadler effect; Jerolmack & Sadler, 2007; Schumer & Jerolmack, 2009) will tend to preferentially obliterate smaller craters (Ref. 20). This is because smaller craters are more likely to be completely removed with the 'Cantor dust' of scale-independent erosion events. This effect is independent of the purely geometric exposure effect discussed in the previous paragraph, although it has the same sign. If the Sadler effect were important for ancient sedimentation on Mars, this would bias our survey towards detecting larger craters. We do not attempt to correct for this bias because we do not know if the Sadler effect was important for ancient sedimentation on Mars. Any correction would lower our paleopressure upper bound, strengthening our conclusions.

We classified one cluster of craters as ancient (in the SE of DTM 1; Fig. S8a). This may be a primary cluster or alternatively might result from dispersal of secondaries in a thicker atmosphere (Popova et al., 2007). It is possible that future work might use ancient crater clusters to set a lower limit on atmospheric paleopressure.

We interpret craters mapped as 'ancient' that lie between the river deposits as being part of the same (buried/embedded) crater population as craters that are overlain by ancient river deposits. If this interpretation is correct, then a histogram of river-crater interaction frequencies from a Monte Carlo trial should be consistent with the measured proportion of craters overlain by ancient river deposits in the measured ancient-crater population. But if our false positive rate is significantly higher away from the river deposits, this would show up as a reduced proportion of river-crater interactions in the measured ancient-crater population relative to that expected by chance as determined by a Monte Carlo trial. For long, parallel river deposits of spacing $W$ and crater diameter < river-deposit width, the fraction of intersections is approximately *D/W*. This is consistent with our mapped populations if we make the approximation *W = A/L* where *A* is DTM area and *L* is channel length. However, the geometry of the real river deposits is more complicated than this idealization (Fig. S8). Therefore, to validate our interpretation, we did the following (typical output shown in Fig. S3):-

(1) Mapped the outlines of all channels within the DTMs (Fig. S8);
(2) Sprinkled random crater populations over the resulting maps, randomly selecting radii from the observed populations and randomizing locations. The number of 'definite' craters and the number of rimmed circular mesas is the same as in the mapped distribution. Craters 100% obscured by channel deposits were removed with replacement;
(3) Counted the number of crater-river interactions for this synthetic population (and the areas of overlap);
(4) Repeated 1,000 times.

We found that the 'definite plus Rimmed Circular Mesas' crater population is in the 56th percentile of the synthetic distribution of crater-river interaction frequencies (Fig. S3). The 'definite' crater population has *more* river-crater interactions than 89% of the synthetic populations, which may indicate a higher likelihood that true embedded craters are relegated to 'candidate' status away from the river deposits. The Rimmed Circular Mesas have a lower interaction frequency than 90% of the random populations, probably because they are locally high-standing so that horizontally-adjacent river deposits have usually been eroded away. This procedure obviously cannot rule out a small contribution of false positives, but in combination with our geologic checklist (Supplementary Table 1) it validates our interpretation that ancient craters mapped as 'definite' between the river deposits do not have a significantly higher false positive rate than ancient craters mapped as 'definite' that are overlain by river deposits.

## 3. Details of data-model comparison.

### 3a. Additional model details.

More details about our forward model of impactor-atmosphere interactions can be found in Williams et al. (2010) and Williams & Pathare (2012). The small-craters technique has been previously applied by Paige et al. (2007) and Kreslavsky (2011) to infer $P$ for relatively recent Martian deposits.

The size distribution of our synthetic impactor populations follows Brown et al. (2002); the initial-velocity distribution follows Davis (1993). Each population contains 3% irons, 29% chondrites, 33% carbonaceous chondrites, 26% cometary objects, and 9% "soft cometary" objects (following Ceplecha et al. 1998) with densities and ablation coefficients $k_{ab}$ also set following Ceplecha et al. 1998. Fragmentation occurs when ram pressure $\varrho_a v^2$ exceeds $M_{str}$, disruption strength. $M_{str}$ is set to 250 kPa; much lower or much higher values would be inconsistent with the observation that more than half of craters observed to form in the current 6 mbar Martian atmosphere are clusters (Daubar et al. 2013). This value of $M_{str}$ is within the range reported for Earth fireballs (Ceplecha et al. 1998), and our conclusions are insensitive to $M_{str}$ variations within the Ceplecha et al. (1998) range. We adopt an impactor entry angle distribution that peaks at 45° (Love and Brownlee, 1991). The ratio of the final rim-to-rim diameter to the transient crater diameter is set to 1.3. The excavation efficiency decreases as $1/(v \sin \theta_i)$ where $\theta_i$ is the impact angle (Pierazzo & Melosh, 2000). We linearly interpolate model output between runs at 0.125, 0.25, 0.5, 1.0, 2.0, 3.0, and 5.0 bars to obtain crater size-frequency distributions as a function of $P$.

We limit the computational cost of the model by only injecting impactors at the top-of-the-atmosphere that are larger than a cutoff diameter $d_c$. Holding $d_c$ constant over the wide range of pressures of interest leads to interminably long runs for high atmospheric pressures. This is because building up a smooth cumulative distribution function of predicted crater diameters (colored lines in Fig. 2) requires hundreds of large impactors, but most CPU time is wasted on detailing the fate of numerous small impactors which have a vanishingly small chance of forming high-velocity craters. Therefore, we set increasing cutoff diameters for increasing atmospheric pressure. These $d_c(P)$ were selected for each $P$ ($P > 0.25$ bar) by progressively decreasing the cutoff diameter from a large value until further reductions did not lead to a significant change in model output crater diameter cumulative distribution function.

**3b. Fitting procedure.**

Atmospheric pressure was found by bayesian fitting of the data to cratering-model output, treating the impacts as a Poisson process (Aharonson et al., 2003; Ch. 6-7 in Wall & Jenkins, 2012).

The power-law slope describing the ratio of large to small impactors is fixed, and the crater density is modeled as a function of atmospheric pressure and an overall impactor frequency. Our procedure is analogous to $\chi$-squared fitting, but it is appropriate for the limit where each bin contains a small number of data.

For each forward model, we ran enough randomized cases to build up a smooth distribution $\lambda = p(D, P)$. When fitting the data to the model, the crater diameters are binned in increments of 1 m. For each of these crater-diameter bins, the probability of obtaining the observed number of craters $Y$ in that size bin given was obtained using Poisson statistics:-

$$p(Y \mid D, P) = \bar{\lambda}^Y \exp(-\bar{\lambda}) / Y!$$

where the overbar corresponds to scaling for the overall number of impacts observed. The overall likelihood of the data given the model is the sum of the logs of the probabilities for each crater-diameter bin (e.g. Ch. 6-7 in Wall & Jenkins, 2012; Aharonson et al., 2003). We separately calculated the best fit paleopressure and statistical error using bootstrapping, obtaining similar results (not shown).

**4. Error analysis and sensitivity tests.**

With ~$10^2$ craters in our sample, the fractional statistical error in our analysis (Supplementary Section 3b) is ~10%. More important are possible systematic errors. In this section, we estimate the individual impact of these possible systematic errors on the conclusions. Because we are reporting an upper limit, we emphasize errors that could raise the upper limit.

- *False positives and false negatives in identifying ancient craters.* In general, orbital imagery of eroding sedimentary-rock units will show a mix of synerosional ("recent") craters and syndepositional (ancient/embedded) craters. Only the ancient craters constrain ancient atmospheric pressure. Because the modern atmosphere of Mars is thin and permits numerous small craters to form, many small craters counted as ancient will be false positives *unless* the base rate of embedded craters is high, or unless the procedure for identifying embedded craters is very accurate (Supplementary Table 1). At the stratigraphic levels mapped in this paper, we observe many craters incompletely overlain by river deposits. Because most of the surface area is not close to the edge of a river deposit (Fig. S8), craters formed in most places would not be overlain by river deposits, or would be completely masked by river deposits (Fig. S8). The observation that many craters are incompletely overlain by river deposits indicates that the base rate of embedded craters is high. Because cratering is random, we expect many embedded craters away from river deposits, and this is consistent with our Monte Carlo results (Supplementary Section 2).

False negatives could in principle bias the results to higher or lower pressures. We documented all "candidate" ancient craters and found that they are smaller on average than the craters used to construct our paleopressure fit (as might be expected from resolution effects). Therefore false negatives do not affect the validity of our upper limit. Having shown that the candidate population does not affect our upper limit, we now provide an extended discussion of this crater population. The 'candidate' exhumed craters – which by definition are not definitely exhumed - may be significantly contaminated by synerosional craters. The regional $N(1)$ count is consistent with a landscape that is currently being sanded down at ~1 μm/yr. Assuming steady state resurfacing with equilibrium between production and obliteration, and ignoring aeolian bedforms, this erosion rate could permit a considerable number of degraded synerosional craters to form in the modern thin atmosphere. However, we do not see many pristine (rayed, blocky, or deep) $D \sim 50m$ craters. It is possible that the balance is made up by 'candidate' exhumed craters that are in fact relatively recent synerosional craters which have lost their rims. The potential for rapid degradation of crater rims in the modern Mars environment is supported qualitatively by evidence of rapid degradation of small craters formed in sedimentary rocks along the Opportunity traverse (Golombek et al., 2010) and rapid degradation of boulders on young fans (Haas et al., 2013). If we are wrong and the candidate exhumed craters are all syndepositional, then our paleopressure upper bound would be lowered by a factor of ~2, strengthening our conclusions.

Channels and channel deposits are identified on the basis of network/tributary structure (Fig. S8), preserved sedimentary structures such as point bars, and double-ridge shape (Williams et al. 2013) in DTM cross-sections. In Aeolis Dorsa, channels are easily distinguished from postdepositional features such as faults.

- *Top-of-atmosphere parameters*. Our model uses a modern (Near Earth Object-like) size-frequency distribution of impactors (Brown et al., 2002), which is relatively rich in small impactors due to faster drift of small asteroids into destabilizing orbital resonances with Jupiter (Strom et al., 2005). This is appropriate for stratigraphic units postdating the Late Heavy Bombardment (see discussion of "Age Control" above); the large rivers on Mars that have been mapped so far were last active significantly after the Late Heavy Bombardment (Fassett & Head, 2008; Hoke & Hynek 2009). If we are wrong and the rivers date from the time of the Late Heavy Bombardment, then the small-impactor-poor impactor size-frequency distribution inferred for the Late Heavy Bombardment by Strom et al. (2005) may be appropriate. In that case, the observation of a large proportion of small impact craters requires an even lower $P$ than reported here, and our paleopressure conclusions are strengthened.

- *Impact parameters and postdepositional modification of impact size and shape*. Crater volume scalings are a physically-motivated fit to experimental data (Holsapple, 1993). Predicted volumes are only accurate to a factor of ~2. Among the parameters in the π-group scaling, the most important parameter sensitivity of the model is to target strength. The strongest rock targets produce decrease in crater size of up to a factor of 2, and a comparable increase in the paleopressure upper bound (Fig. S4b), relative to our preferred rock-mass strength of 65 kPa (Refs. 21, 22; see also http://keith.aa.washington.edu/craterdata/scaling/theory.pdf). Our main argument against adopting strong-rock rock-mass-strength for our model is geological – because of the observed fine layering and high density of river deposits (Refs. 11, 20; Fig. S1), the simplest interpretation of geological units "F1" and "F2" is that they are fluvial/alluvial or other weak sedimentary deposits, analogous to terrestrial desert alluvium. Desert alluvium has been thoroughly characterized through Nevada Test Site explosions of comparable energy to the small natural high-velocity impact craters used in this paper, and an empirical rock-mass strength of ~65 kPa is inferred. This is the value that we use in this paper. Crucially, the *present-day* outcrop strength of the Aeolis Dorsa deposits is irrelevant, because embedded craters formed early in the history of the deposits and the timing of any compaction or cementation is unknown. Model output is not very sensitive to the details of how fragmentation is parameterized ($\lesssim$10%; Fig. S4a), nor to target density ($\lesssim$25% for range 1500-2500 kg/m$^3$; Fig. S4c), nor to reasonable variations in the mix of impactor strengths and densities (e.g., the stone:iron ratio; not shown). Setting $\mu$ = 0.55 (as opposed to our adopted value of $\mu$ = 0.41; Methods) is reasonable if ice, groundwater, or diagenetic cements filled the pore spaces of the target material. For fixed target strength, this increases crater diameters, typically by a factor of ~5/3 (Fig. S4b). If $\mu$ = 0.55 then (holding all other parameters constant) the observed small impact craters would correspond to even smaller impactors surviving passage through the paleoatmosphere. This would strengthen our conclusions.

- As discussed in the main text, *erosion may modify craters*. Our main safeguard against this source of error is to fit the circles defining the crater diameters only to parts of the crater edge which are well-preserved. A supplementary safeguard is to expand (or contract) the resulting circles until they enclose only two (or enclose all *except* two) of the hand-picked points on the crater rim. We then define the annulus enclosed by these minimal and maximal circles as a 'preservation-error annulus.' This accounts for possible erosional modification of crater shapes, assumed to be initially close to circular (Melosh, 1989). The full width of the annulus was (13±6)% of nominal diameter for definite embedded craters and (16±7)% of nominal diameter for RCMs. We found no significant difference between total errors (from resampling) including random sampling of radii from within the preservation-error annulus as opposed to total resampling errors excluding this effect.

- *Errors in elevation* propagate to errors in the final Mars paleo-atmospheric pressure estimate because they affect the hydrostatic correction of pressure to zero elevation (i.e. to the Mars datum). In this context, the intrinsic error of the DTMs is negligible (<<100 m), because they are controlled to the Mars Orbiter Laser Altimeter dataset which has a radial precision of ~1m (Smith et al., 2001). The elevation range of the studied craters is ~0.1 km (~1% of an atmospheric scale height), which is also negligible. Even if postdepositional tectonic uplift/subsidence of the studied terrain had an amplitude of 1 km (which is unlikely), this would introduce a systematic error of only ~10%.

In summary, the error in our upper limit on $P$ is set primarily by uncertainty in the effective rock-mass strength of the target at the time of impact. Our chosen strength value follows from our geologic interpretation of the target materials; if our geologic interpretation is correct, then the $P$ error due to strength uncertainty is <50%. If our geologic interpretation is incorrect, then this could introduce an error of (at most) a factor of 2, but this is counterbalanced to some degree by the possibility that $\mu$ was higher than the value we have chosen here. In the future, small-scale lab experiments, crater-counts of geologic materials of similar age but different strengths (e.g. Ref. 21), and ground-truth from rover observations could better constrain these errors.

## 5. DTM extraction procedure.

The procedure used for DTM extraction follows that of Ref. 10 and uses the NGATE algorithm (Zhang, 2006) and SOCET SET software. The HiRISE images making up the PSP_007474_1745/ESP_024497_1745 steropair have emission angles of 4.5° and 30° respectively, and map scales of 25 cm/pixel and 50 cm/pixel respectively. The coarser image (ESP_024497_1745 in this case) determines the optimal spatial resolution for the topographic extraction, so we derived a 2.5 m/post DTM for this pair (DTM1). MOLA PEDRs were used as ground control points, with vertical accuracy set to 10 m, as the area contains mostly flat smooth features, for which it is difficult to link PEDR shots to surface features observed at HiRISE scale. In addition, we generated our own gridded MOLA DTM (from PEDR), which we used as a seed for extraction. The process for DTM2 was very similar (emission angles 2° and 18°; map scales of 50 cm/pixel for both images).

We used several metrics for DTM validation and quality assessment. These included LE90 (Linear Error of 90%). This value is automatically computed (by the SOCET SET photogrammetry software) as the error in elevation of one point with respect to another point within the DTM at 90% probability. In DTM1, the mean LE90 is 1.07 m and when correlation had succeeded, the highest value is 3 m. These values should be compared with the theoretical limit on vertical precision using the standard photogrammetry equation (Ref. 10):

$$EP = r\, s\, /\, (b/h)$$

where $EP$ is the expected vertical precision, $r$ is the accuracy with which features can be matched (i.e., $r = 0.3$), $s$ the ground sample distance (i.e., $s = 50$ cm), and the $b/h$ ratio describes the convergence geometry of the stereopair (i.e., $b/h \sim 0.5$). These values give $EP \sim 0.3$ m. As a test, the shaded relief was compared to the orthophoto using the same illumination geometry over a constant albedo area (Figs. S5, S6). We also compared cross-sections over both the HiRISE image and the shaded relief computed from the DTM. A good match was obtained.

**Supplementary References.**

Williams, R.M.E., et al., Variability in martian sinuous ridge form: Case study of Aeolis Serpens in the Aeolis Dorsa, Mars, and insight from the Mirackina paleoriver, South Australia, *Icarus* **225**, 308-324 (2013).

Zhang, B., Towards a higher level of automation in softcopy photogrammetry: NGATE and LIDAR processing in SOCET SET®, paper presented at GeoCue Corporation 2nd Annual Technical Exchange Conference, Nashville, Tenn. (2006).

Zimbelman, J.R., & Scheidt, S.P., Hesperian age for Western Medusae Fossae Formation, Mars, *Science* **336**, 1683 (2012).
31Williams, R.M.E., et al., Variability in martian sinuous ridge form: Case study of Aeolis Serpens in the Aeolis Dorsa, Mars, and insight from the Mirackina paleoriver, South Australia, *Icarus* **225**, 308-324 (2013).

Zhang, B., Towards a higher level of automation in softcopy photogrammetry: NGATE and LIDAR processing in SOCET SET®, paper presented at GeoCue Corporation 2nd Annual Technical Exchange Conference, Nashville, Tenn. (2006).

Zimbelman, J.R., & Scheidt, S.P., Hesperian age for Western Medusae Fossae Formation, Mars, *Science* **336**, 1683 (2012).
31

**Supplementary Table 1: Checklist for identifying ancient craters.**

Figure S7 shows examples of applying the checklist, and Figure S8 shows the crater maps resulting from applying the checklist.

---

**Checklist for accepting ancient craters**

*Must be an impact structure that is embedded within the stratigraphy.*

- Crater, or crater rim (if preserved), or ejecta (if preserved) are crosscut by fluvial deposits → accept
- Crater, or crater rim (if preserved), or ejecta (if preserved) are crosscut by fluvial channels → accept
- Crater partly overlain by sediments topographically, stratigraphically or texturally continuous with surrounding layered sediments → accept
- Crater forms a rimmed circular mesa
- Crater forms a rimmed circular mesa with flat or inward-dipping strata inside the rim; these strata need not be continuous with sediment outside (and usually are not)

*Other checks:*

- At same or similar level and spatially adjacent to an ancient crater; has the same preservation style (e.g., layered circular mesa) as that ancient crater
- Crater is close to circular (ellipticity < 1.15)

- Rim or edge preserved topographically in DTM over at least 180° of arc (does not have to be continuous)

*or*

- Crater appears to be concave-up in anaglyph

*if neither:*

- Reject.

---

**Ensemble checks:**

- Is the same preservation style of craters found beyond the mapped background geologic unit in this geologic region? (If so, could be a younger mantling unit: reject all craters)
- Are the ellipticities aligned?
- Is the distribution of crater centers random in space?
- Are any clusters of craters restricted to a particular stratigraphic level or a particular geologic unit? (If so, suspect soft-sediment deformation).

**Checklist for rejecting ancient craters: rejects override accepts**
*Either not clearly an impact structure, or not embedded within stratigraphy*

- Rim preserved mostly (>2/3) intact, and rim ellipticity > 1.5 → immediate reject
- Crater (and ejecta, if visible) are not superposed by anything other than active/recently active bedforms → immediate reject
- Rays visible → immediate reject
- Crater could be a prolongation of nearby soft-sediment deformation texture consisting of cells with upcurled edges ('spatulate' soft-sediment deformation).
- (For circular mesas) The height of the mesa exceeds the radius of the flat top or rim by >1.5 (risk of being a rootless cone or explosion pit analogous to von Braun/Goddard at the Spirit field site in Gusev crater).
- There is a rim visible around all or most of the top of the structure, but the elevation of the rim is much lower on one side of the structure (immediate reject; suggestive of volcanism or soft-sediment deformation)

**Ensemble level checks for circular mesas -** Is there a connection between the relief of the mesa and the diameter of the depression on top? if yes, argues for explosive cone rather than eroded/exhumed impact crater.

**Supplementary Figures.**

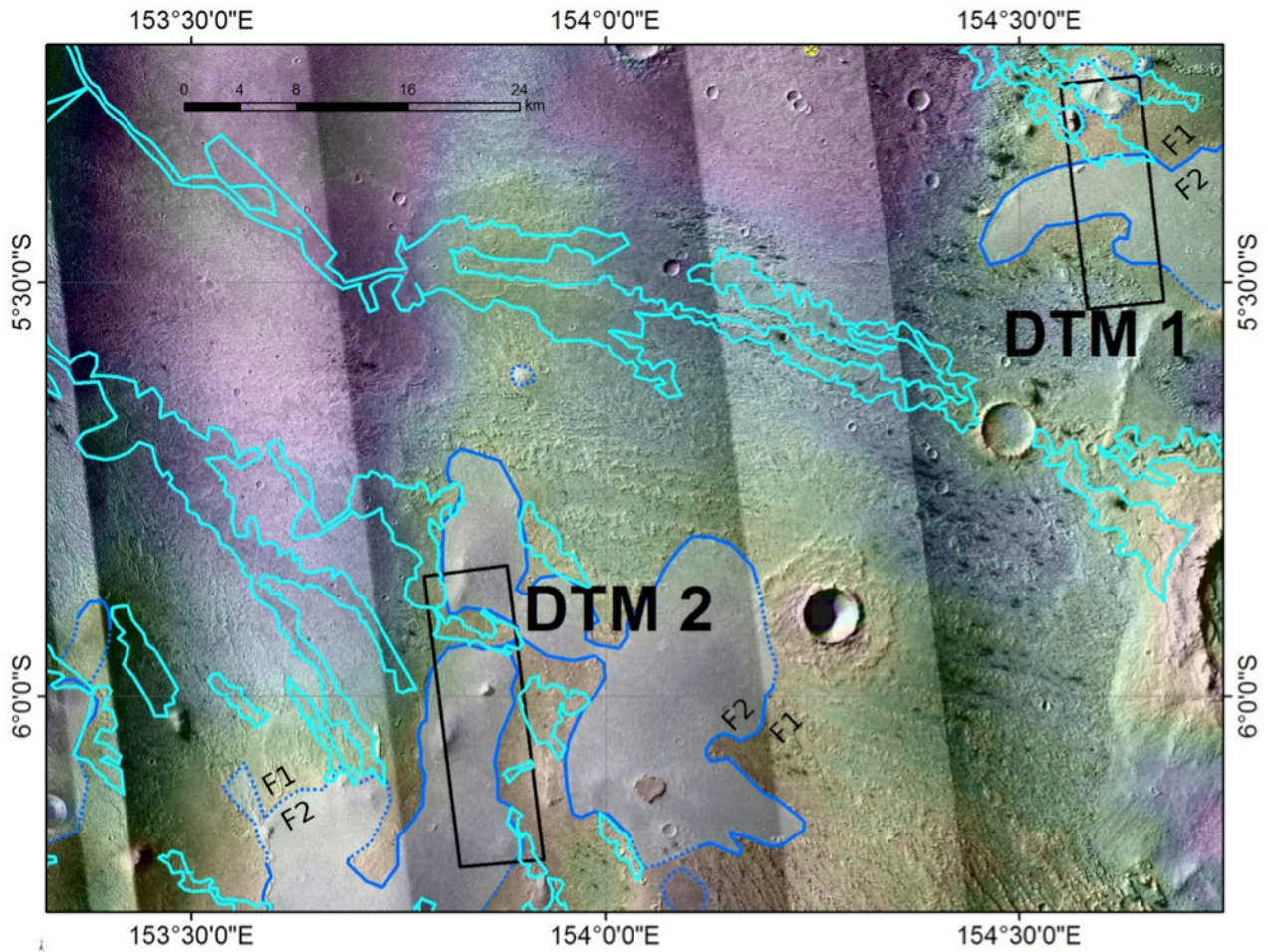

**Figure S1.** Geologic context for this study. Topographically lower fluvial unit ("F1", no tint) contains large meander belts (cyan outlines). Topographically higher fluvial unit ("F2", white tint) contains many river deposits but lacks large meander belts. F1/F2 contact is shown as a solid blue line where mapped with high confidence, and as a dotted blue line where inferred. Background color is cued to MOLA topography (elevation range ~ 500m). Background image is CTX mosaic; the western rim of Kalba crater is visible at right. DTMs were constructed from HiRISE images PSP_007474_1745/ESP_024497_1745 (DTM 1) and ESP_017548_1740/ESP_019104_1740 (DTM 2). DTM1 area is 108 km$^2$; DTM2 area is 86 km$^2$. See Fig. S8 for details of DTMs.

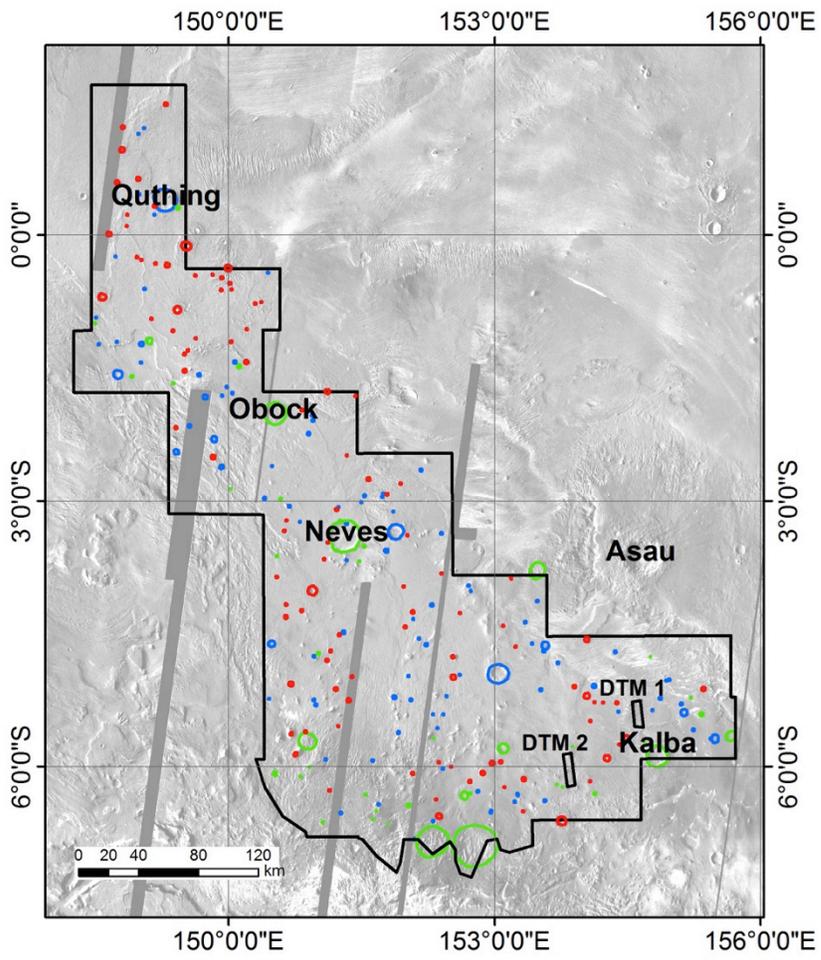

a)

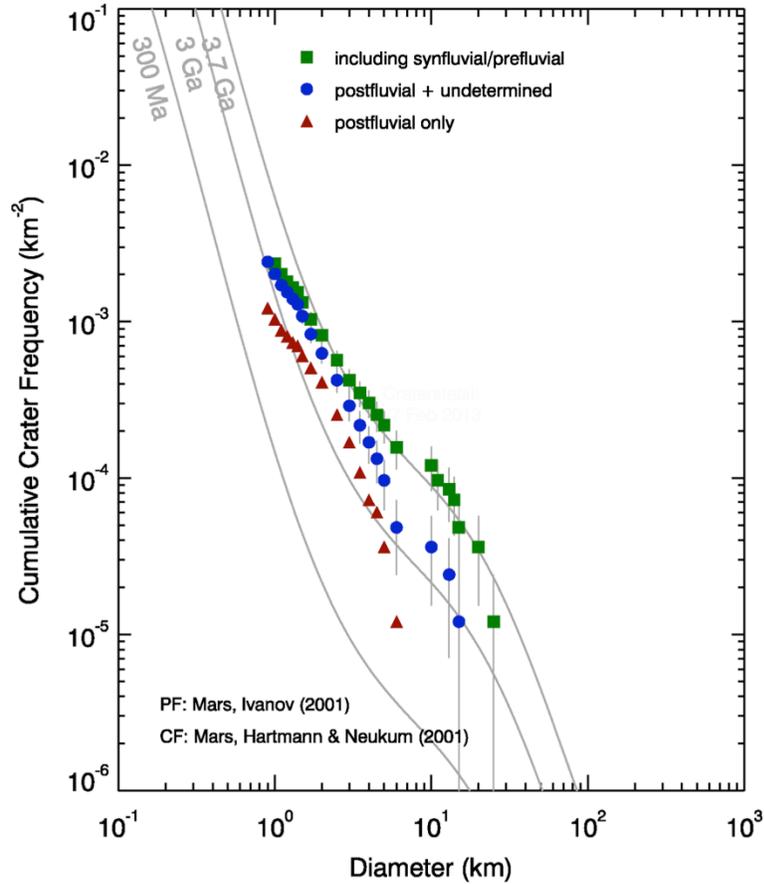

b)

**Figure S2.** Chronologic context for this study. a) Locations of all craters >1km diameter. Red corresponds to craters that are postfluvial based on local crosscutting relationships; blue corresponds to craters with an undetermined crosscutting relationship to nearby rivers (these are interpreted to be mostly postfluvial on the basis of crater morphology); and green corresponds to synfluvial/prefluvial craters. Black polygon corresponds to perimeter of count area (8.3 x $10^4$ $km^2$). Background is THEMIS VIS mosaic. b) Cumulative crater size-frequency distributions plotted using `craterstats2` (Michael & Neukum 2010). Error bars show 1σ statistical error. Red: postfluvial craters only. Nominal age considering only crater diameters >2 km is 3.44 (+0.06/–0.10) Ga. Blue: additionally including "undetermined" craters. Nominal age considering only crater diameters >2 km is 3.61 (+0.03/–0.04) Ga. We consider this a lower bound on the true age of Aeolis Dorsa rivers (see text). Green: additionally including prefluvial/synfluvial craters. Nominal age considering only crater diameters >2 km is 3.71 (+0.02/–0.03) Ga.

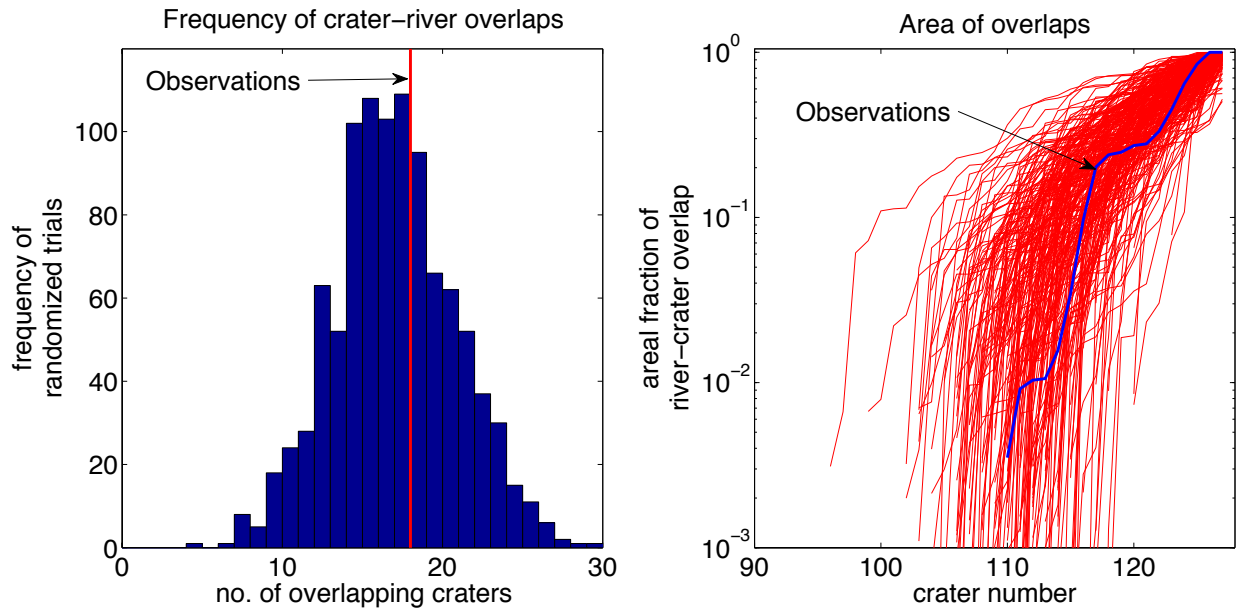

**Figure S3.** Comparison of crater-river interactions in the observed population to an ensemble of synthetic crater populations with the same size-frequency distribution. For assumptions, see text. Left panel: Frequency of crater-river overlaps for 1,000 synthetic crater populations (observations shown by vertical red line). Right panel: Crosscut test comparing observed crater-river interaction areas to an ensemble of 1,000 synthetic crater-populations. Ordinate corresponds to fractional area of overlap for each crater – for legibility, only every fourth synthetic population is shown. Craters are sorted by fractional overlap – the majority of craters in the synthetic and observed populations have zero overlap. Observations are shown by thick blue line.

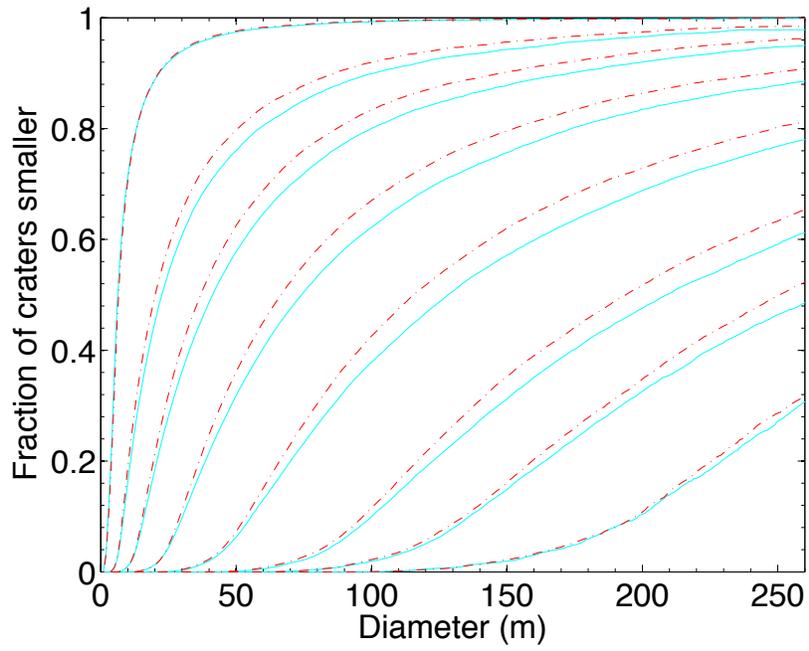

(a)

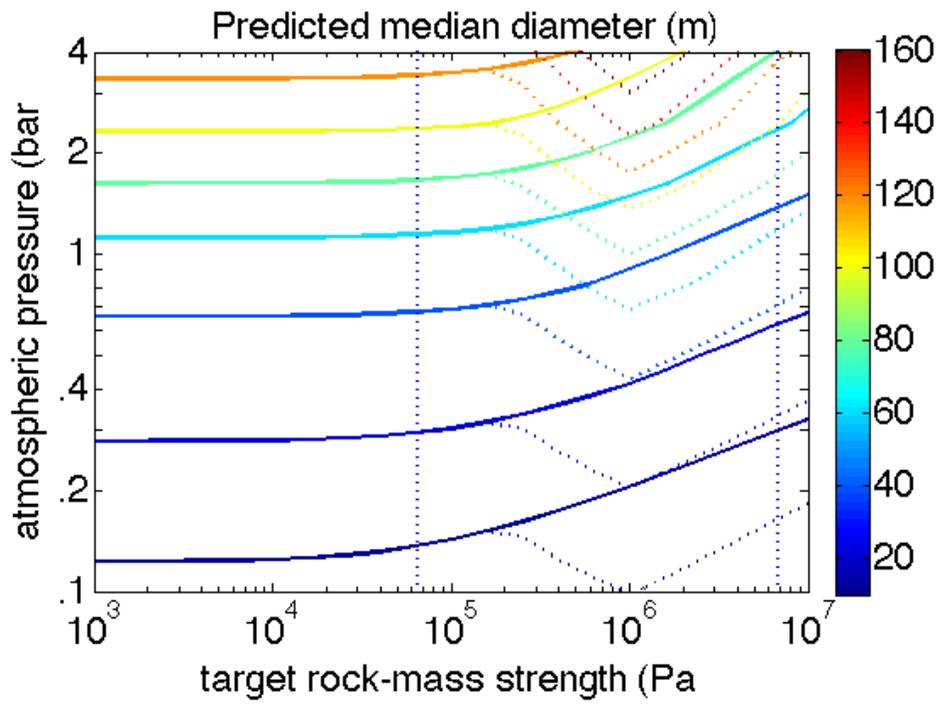

(b)

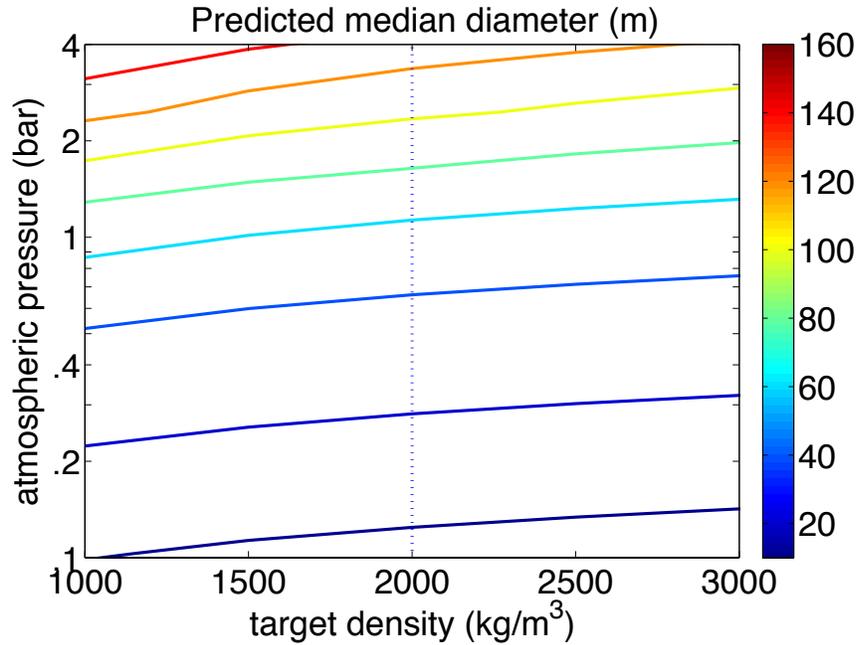

(c)

**Figure S4.** Sensitivity tests. (a) Fragmentation parameterization: cyan solid lines show crater sizes ignoring the last fragmentation event; red dashed lines show "effective" size of impact combining all fragments into one "effective" cluster. From left to right, pressures are for 6 mbar, 125 mbar, 250 mbar, 500 mbar, 1 bar, 2 bar, 3 bar and 5bar (assuming impacts at 0m elevation). (b) Sensitivity to target rock-mass strength (using $\pi$-group scaling; Refs. 14, 22). Contours drawn at median crater size of 10 m, 20 m, 40 m, and then at 20 m intervals until 160 m. Left vertical dashed line (65 kPa) is strength inferred for desert alluvium (Ref. 14), which is appropriate to our geologic setting. Right vertical dashed line (6.9 MPa) is "hard rocks" value used by Ref. 22 (their Figure 7). Solid lines correspond to constant $\mu = 0.41$; colored dashed lines show effect of log-linear ramp of $\mu$ from 0.41 at 200 kPa to 0.55 at 1 MPa and constant thereafter (`http://keith.aa.washington.edu/craterdata/scaling/theory.pdf`). If the Aeolis Dorsa sediments had "hard rock"-like strength and $\mu = 0.41$ at the time the craters formed, then our upper limit is significantly relaxed. (c) Sensitivity to target density (using $\pi$-group scaling): Contours drawn at median crater size of 10 m, 20 m, 40 m, and then at 20 m intervals until 140 m. Vertical dashed line is our preferred value (2000 kg/m$^3$); a reasonable range is 1500 – 2500 kg/m$^3$, for which inferred-paleopressure variations are modest.

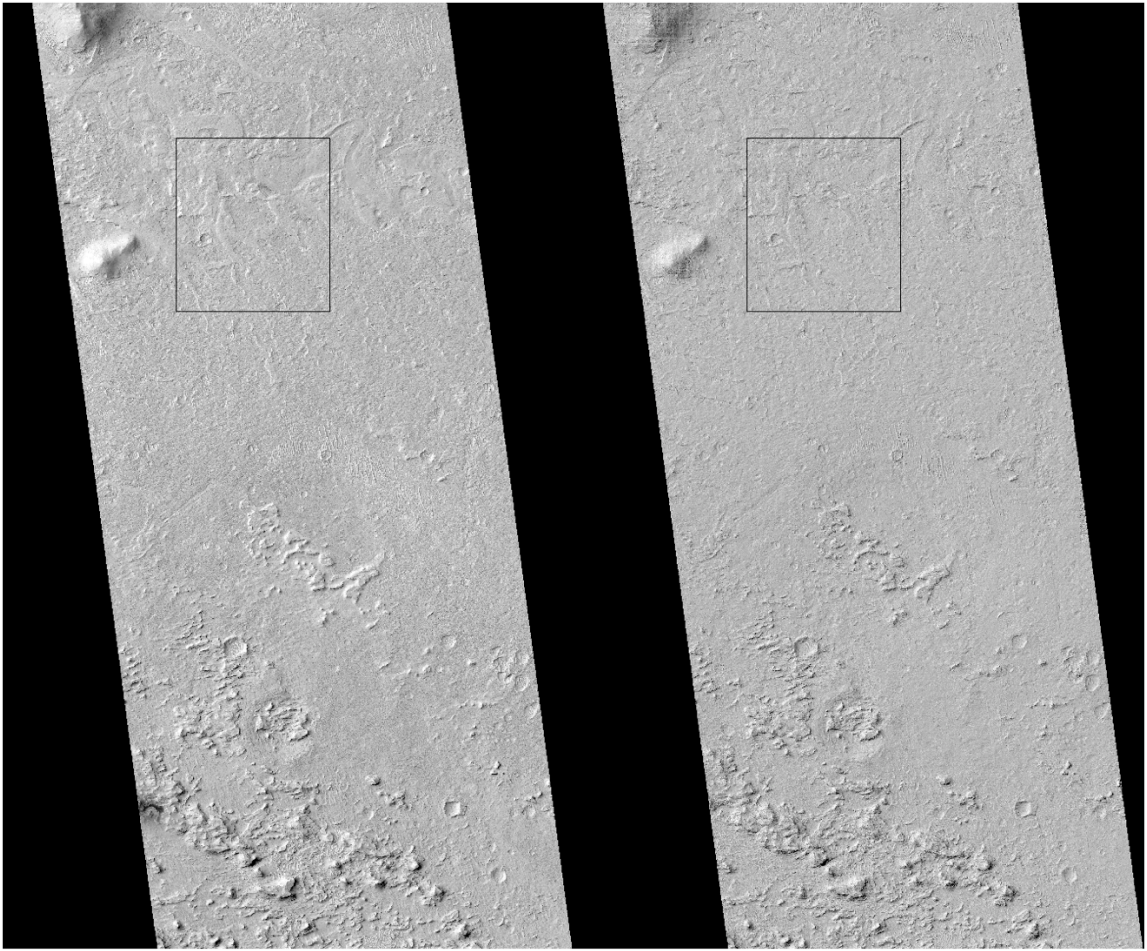

**Figure S5.** PSP_007474_1745 image on left, shaded relief of corresponding DTM (DTM1, PSP_007474_1745/ESP_024497_1745) on right illuminated using the same illumination geometry as the image. Black box shows region highlighted in Fig. S6.

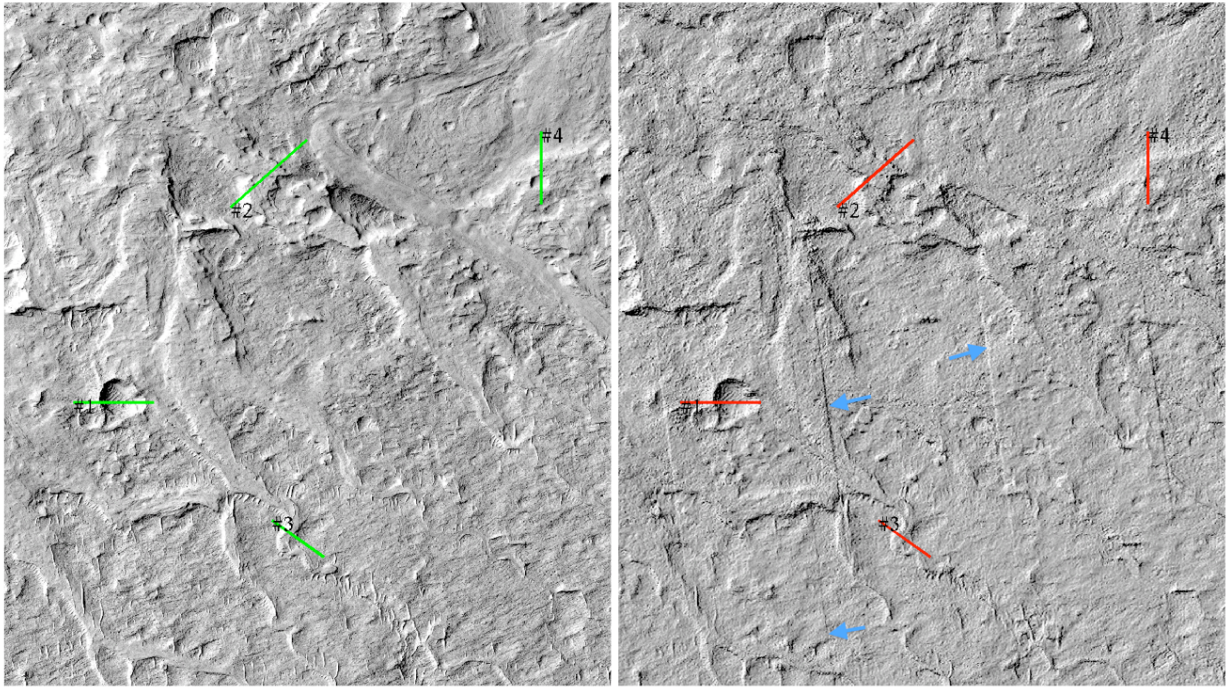

**Figure S6.** Comparison between HiRISE image and a shaded relief of the corresponding stereo DTM using the same illumination geometry. Left panel: PSP_007474_1745 image (25cm/pixel). Right panel: shaded relief from the stereo extraction. Seams at the boundaries between HiRISE CCDs are visible in the DTM (blue arrows on right panel). Their obvious presence makes it possible to take them into account in any measurement. Red and green profiles highlight points of agreement.

**Figure S7.** Examples of application of the checklist in Supplementary Table 1 (anaglyphs not shown). DEF = definite embedded crater; RCM = rimmed circular mesa; CAND = candidate ancient crater (excluded from paleopressure calculations). Key to sketch interpretations: *c – crater or crater fill; cand – candidate ancient crater; ch – channel or channel-fill material; cr – crater rim material; fl – fluvial deposits not part of an integrated channel; ifm –interfluve material (unknown origin; simplest interpretation is fluvial overbank material); rcm – rimmed circular mesa.*

| Type | Orthophoto | Orthophoto + DTM | Sketch interpretation | Notes |
|---|---|---|---|---|
| DEF | 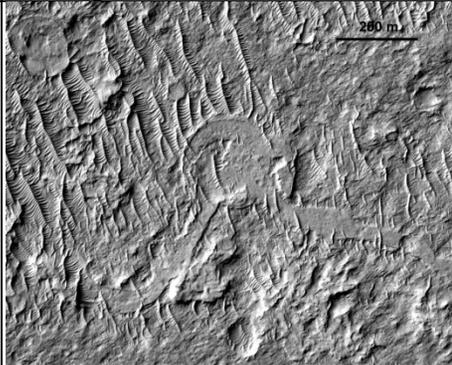 | 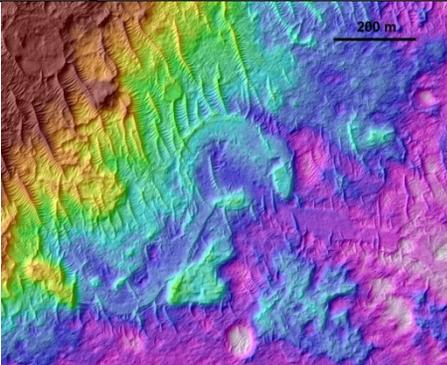 | 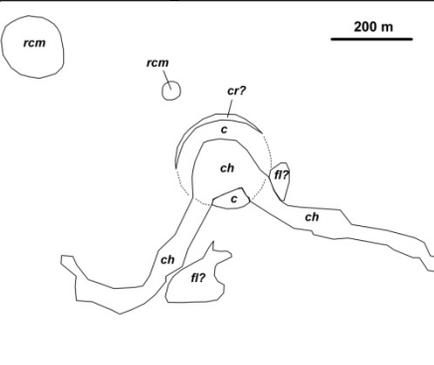 | Crater is crosscut by fluvial deposits that are topographically and texturally continuous with those outside crater. Crater is close to circular. Rim or edge is preserved (discontinuously) over more than 180° of arc. → DEF. ESP_017548_1740. See also Figure 1 for additional examples of definite embedded craters. This crater is entry #8 from the Supplementary Table of Ref. 20. |
| RCM | 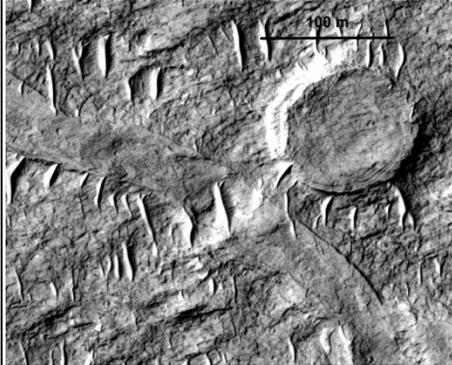 | 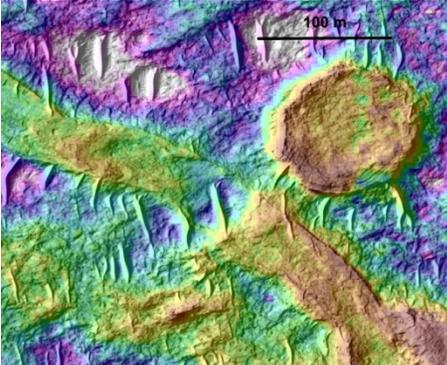 | 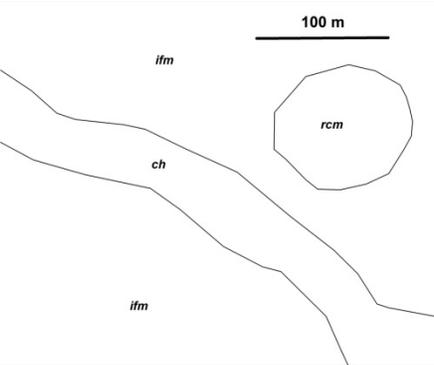 | Crater forms a rimmed mesa. Ellipticity is < 1.15. Rim is preserved (based on DTM and image shading) over more than 180° of arc. Crater appears concave-up in anaglyph and in DTM. No evidence for rays, ejecta, or nearby soft-sediment deformation of similar style. Elevation of mesa ~4m, much less than mesa diameter. → RCM. PSP_07474_1745. |
| CAND | 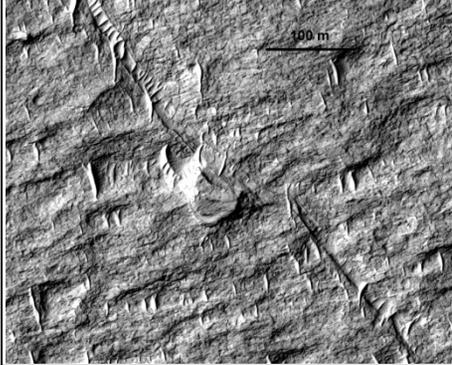 | 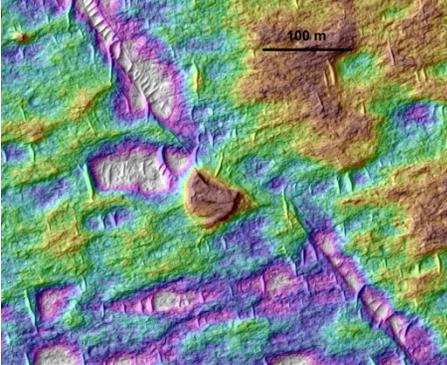 | 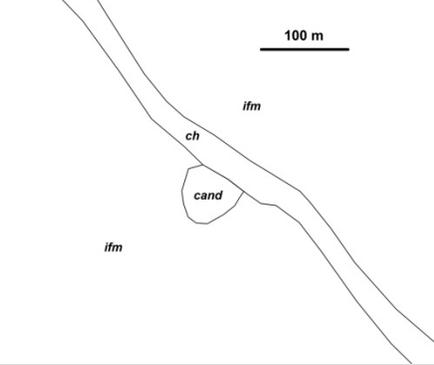 | Raised circular structure is truncated by channel, interpreted as fluvial channel based partly on network structure not visible in this subframe. Subtle rim may be present, however structure is convex-up overall. Origin is unclear: one possible alternative to impact is preferential erosion around the margins of spatulate soft-sediment-deformation. → CAND. PSP_07474_1745. |

**Figure S8.** Maps showing locations of:- definite ancient craters (green); rimmed circular mesas (orange); candidate ancient craters (red - excluded from paleopressure calculations); channels and channel belts (gray shading). In most cases crater rims are only partially preserved.

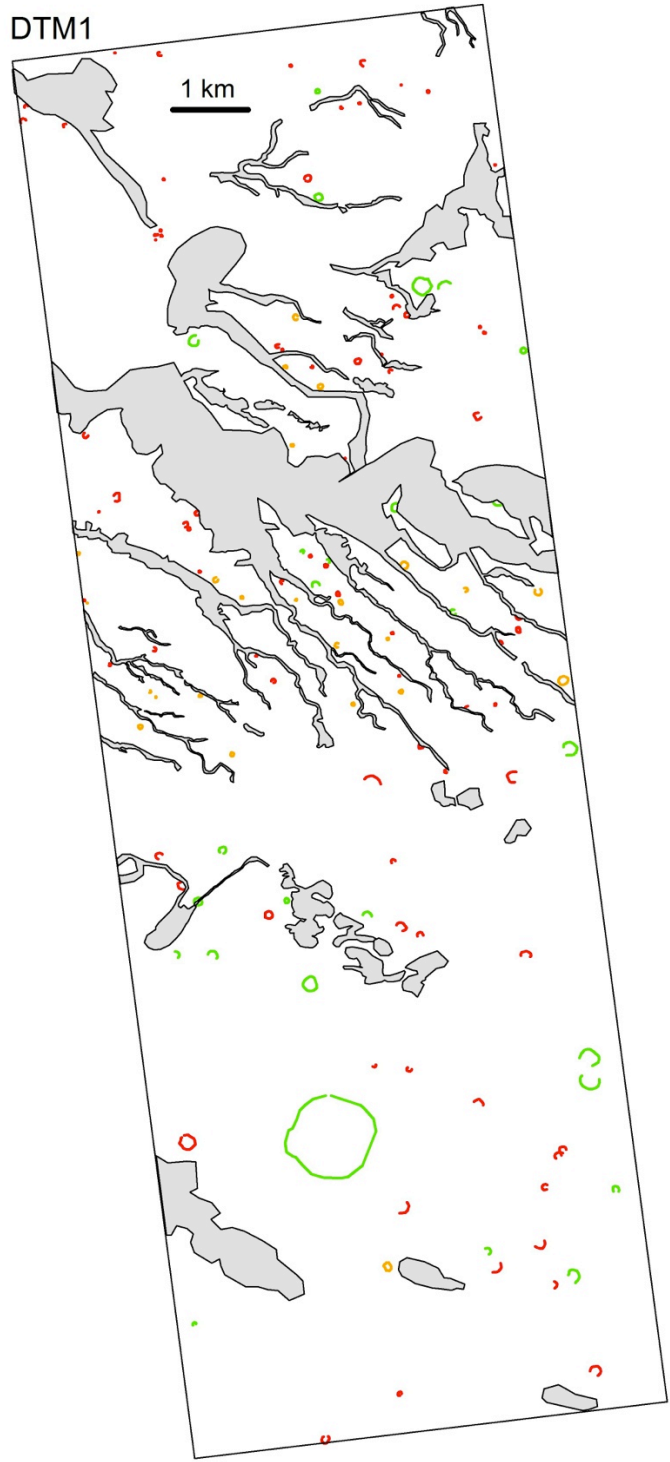

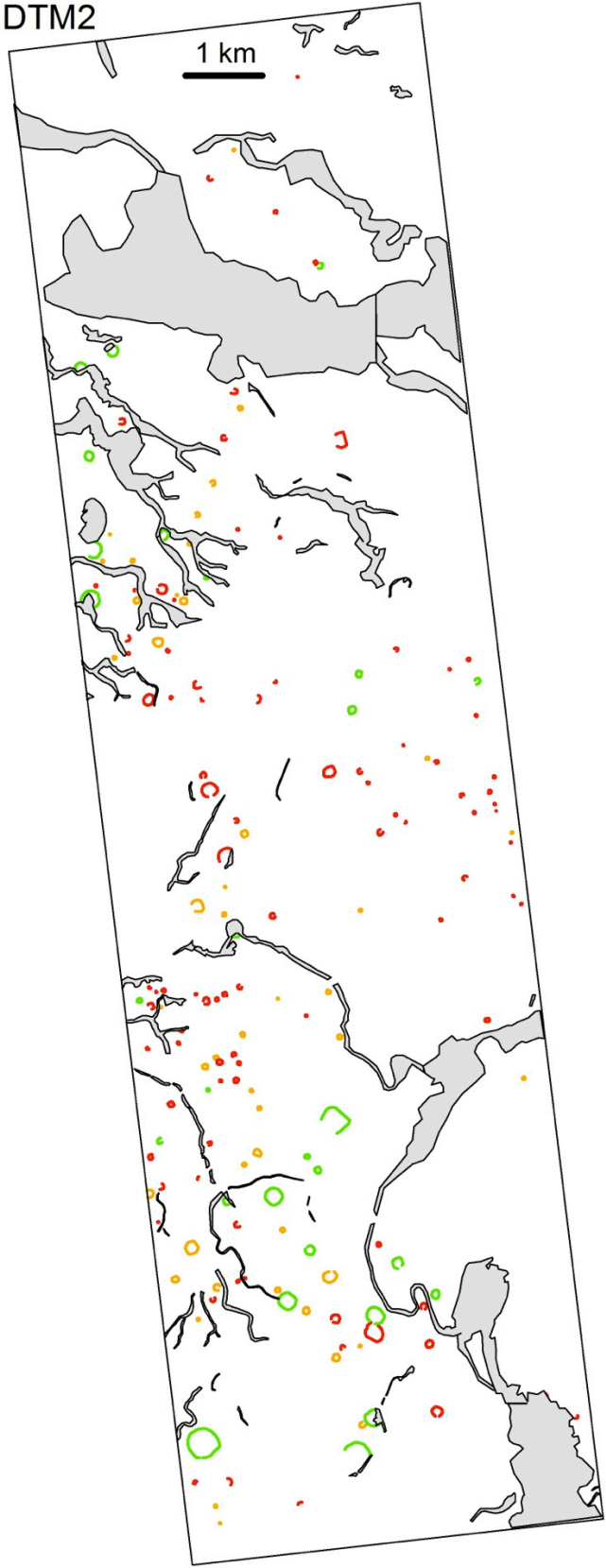